\documentclass[letterpaper,11pt]{article}

 \title{A copula based approach for electoral quick counts}
 \author{Arturo Erdely\thanks{Personal website https://sites.google.com/site/arturoerdely}}
 \date{\small{Facultad de Estudios Superiores Acatl\'an \\
              Universidad Nacional Aut\'onoma de M\'exico \\
							\texttt{arturo.erdely@comunidad.unam.mx}\\}}
 
 \usepackage{amsthm}
 \usepackage{amssymb}
 \usepackage{amsmath}
 \usepackage{enumerate}
 \usepackage{graphicx}

\newcommand{\prob}{\mathbb{P}}
\newcommand{\esper}{\mathbb{E}}

\newcommand{\vari}{\mathbb{V}}
\newcommand{\cov}{\mathbb{C}\text{ov}}
\newcommand{\indic}{\textbf{\textsf{\large{1}}}}

\theoremstyle{plain}

\theoremstyle{definition}

\theoremstyle{remark}

\begin{document}
 
\maketitle

\begin{abstract}
  \noindent An electoral quick count is a statistical procedure whose main objective is to obtain a relatively small but representative sample of all the polling stations in a certain election, and to measure the uncertainty about the final result before the total count of votes. A stratified sampling design is commonly preferred to reduce estimation variability. The present work shows that dependence among strata and among candidates should be taken into consideration for statistical inferences therein, and a copula based model is proposed and applied to Mexico's 2006, 2012, and 2018 presidential elections data.
\end{abstract}


\bigskip

\noindent\underline{INDEX}\par\medskip
1. Introduction\par
2. Stratified sampling design\par
3. Interval estimation under dependent strata\par
4. A copula based model\par
5. Mexico's 2006, 2012, and 2018 presidential elections\par
6. Final remarks\par

\section{Introduction}

\noindent Accordingly to Lindley (2000) the main objective of the discipline known as \textit{Statistics} is the study, quantification and combination of uncertainties, mainly based on sample data:
\begin{quote}
  \textsl{``\ldots the statistician's role is to assist workers in other fields, the clients, who encounter uncertainty in their work. In practice, there is a restriction in that statistics is ordinarily associated with data; and it is the link between the uncertainty, or variability, in the data and that in the topic itself that has occupied statisticians [\ldots] A scientific approach would mean the \textbf{measurement of uncertainty}; for, to follow Kelvin, it is only by associating numbers with any scientific concept that the concept can be properly understood.''}
\end{quote} 

\noindent In the particular case of an electoral process where a single person is to be elected (president, governor, etc.) by direct vote of the people, the main uncertainties of interest to be measured, based on sample data and before the total vote count is finished, are the following:
\begin{itemize}
  \item Who will be the winner?
	\item Percentage of votes obtained by each candidate.
\end{itemize}

\smallskip

\noindent An \textit{electoral quick count} is a statistical procedure whose main objective is to obtain a relatively small but representative sample of all the polling stations in a certain election to make inferences about the above two questions. The uncertainty about the winner may be measured as a \textit{probability of victory} of the leading candidate, while the percentage of votes may be inferred through \textit{point and interval estimations}. A \textit{stratified sampling design} is commonly preferred to reduce estimation variability.

\bigskip

\noindent In the following sections, after introducing appropriate notation and the stratified sampling design, the need of considering dependence among strata is discussed for the case of interval estimation, and the need of considering dependence among candidates is discussed to estimate the probability of victory of the leading candidate, including a copula based model which is applied to Mexico's 2006, 2012, and 2018 presidential elections data, and compared to the model proposed by Mendoza and Nieto-Barajas (2016).

\section{Stratified sampling design}

\noindent The notation used in Mendoza and Nieto-Barajas (2016) will be mainly adopted or adapted, since a comparison to their proposed model will be made later. The available \textit{sample frame} to be considered is a total of $\,K\,$ polling stations for a total of $\,n\,$ potential voters, who may cast their vote in favor of:
\begin{itemize}
  \item a registered candidate,
	\item non-registered candidates,
	\item nobody (null vote),
	\item abstention (no-show at polling station).
\end{itemize}

\smallskip

\noindent The above voting possibilities will be labeled as $1,2,\ldots,J$ where $J-2$ will always be for non-registered candidates, $J-1$ for null votes and $J$ for abstentions, so that categories $1,\ldots,J-3$ will always be for registered candidates, implying $J\geq 5$ (at least two registered candidates). In the following, when reference is made to a candidate $j\in\{1,\ldots,J\}$ this should be understood in a broader sense, considering also as ``candidates'' categories such as non-registered candidates, null votes and abstentions.

\bigskip

\noindent It will be considered that the total of $K$ polling stations are distributed into $N$ non-overlapping subsets or \textit{strata} (electoral districts, for example). Let $K_i$ be the number of polling stations in stratum $i\in\{1,\ldots,N\}$ and let $n_i$ be the number of potential voters in stratum $i$ so that $\sum_{i=1}^N n_i=n.$ When polling stations close, all the votes at each polling station are necessarily classified into categories $1,\ldots,J-1.$ Let the random variable $X_{i,j}^{(k)}$ be the number of votes for candidate $j\in\{1,\ldots,J\}$ in stratum $i\in\{1,\ldots,N\}$ and polling station $k\in\{1,\ldots,K_i\}.$ Also, let the random variables $X_{i,j}=\sum_{k=1}^{K_i}X_{i,j}^{(k)}$ be the total number of votes for candidate $j$ in stratum $i,$ and let the random variables $\Theta_{i,j}=X_{i,j}/n_i$ be the proportion of votes in favor of candidate $j$ in stratum $i.$ Then it is immediate that $\sum_{j=1}^J X_{i,j}=n_i\,$ and $\,\sum_{j=1}^J\Theta_{i,j}=1$ for each stratum $i.$

\bigskip

\noindent With the above it is possible to define random variables $X_j$ as the total number of votes for candidate $j$ and, consequently, $X_j=\sum_{i=1}^N X_{i,j}=\sum_{i=1}^N n_i\Theta_{i,j}\,.$ The overall proportion of votes for each candidate $j$ is then given by the random variables:
\begin{equation}\label{eq:propVotos}
  \Theta_j \,\,=\,\, \frac{X_j}{n} \,\,=\,\, \sum_{i\,=\,1}^N\frac{n_i}{n}\,\Theta_{i,j}\,,\qquad j\in\{1,\ldots,J\}
\end{equation}
from where it is immediate to verify that $\sum_{j=1}^J X_j=n\,$ and, consequently, $\,\sum_{j=1}^J\Theta_j=1.$ The proportion of \textit{effective participation} in the election is calculated as the complement of abstentionism, that is $1-\Theta_J\,.$

\bigskip

\noindent It is considered the usual case where a \textit{stratified sampling design} is applied, with simple random sampling within strata, proportional to the number of potential voters in each stratum. That is, if in stratum $i$ there are $K_i$ polling stations, among these a subset of $c_i\ll K_i$ polling stations will be drawn at random, so that if $c=\sum_{i=1}^N c_i$ represents the total number of polling stations in the stratified sample then $c_i\approx \frac{n_i}{n}\,c\,.$ The total sample size $\,c\,$ depends on a desired \textit{precision} of the estimators to be used, which may be controlled by setting a \textit{margin of error} $\varepsilon>0$ and a \textit{confidence level} $100\alpha\%$ (where $0<\alpha<1$). If $\theta_j$ is the unknown proportion of votes for candidate $j$ and $\widehat{\theta}_j$ is a point estimator of $\theta_j$ then the total sample size $\,c\,$ could be chosen such that:
\begin{equation}\label{precision}
  \mathbb{P}(\,|\theta_j\,-\,\widehat{\theta}_j|\leq\varepsilon\,)\,=\,\alpha\,,
\end{equation}
but for each $j$ the distribution of the point estimator $\widehat{\theta}_j$ depends on unknown population parameters, so data from a previous or similar election could be used, and even proceeding this way, for each $j$ a different value for $\,c\,$ may be obtained to achieve (\ref{precision}), therefore the largest value for $\,c\,$ could be used to (approximately) ensure that for all $j\in\{1,\ldots,J-1\}$ it is possible to achieve at least $100\alpha\%$ confidence, that is $\mathbb{P}(|\theta_j-\widehat{\theta}_j|\leq\varepsilon)\geq\alpha.$ The inequality in (\ref{precision}) is equivalent to $\,\widehat{\theta}_j-\varepsilon\,\leq\,\theta_j\,\leq\,\widehat{\theta}_j+\varepsilon,$ which allows the interpretation that the unknown value $\theta_j$ will belong to the interval $\,[\,\widehat{\theta}_j-\varepsilon\,,\,\widehat{\theta}_j+\varepsilon\,]\,$ with (at least) a $100\alpha\%$ confidence level. As a consequence of Result 3.7.2 in S\"{a}rndal \textit{et al.}(1992) a point estimator for $\theta_j$ may be, for example:
\begin{equation}\label{estimadorderazon}
  \widehat{\theta}_j\,=\,\frac{\sum_{i=1}^N K_i\hspace{0.3mm}\overline{y}_{i,j}}{\sum_{\ell=1}^{J}\sum_{i=1}^N K_i\hspace{0.3mm}\overline{y}_{i,\ell}}\,,\qquad j\in\{1,\ldots,J\},
\end{equation}
where $\,\overline{y}_{i,j}\,$ is the average number of votes obtained by candidate $j$ in the sample of stratum $i.$

\section{Interval estimation under dependent strata}

\noindent The total proportion of votes in favor of candidate $j$ in (\ref{eq:propVotos}) is a convex linear combination $\,\Theta_j=\sum_{i=1}^N\beta_i\Theta_{i,j}\,$ where $\,\beta_i=n_i/n>0\,$ and consequently $\,\sum_{i=1}^N\beta_i=1.$ Let $\,\mu_{i,j}:=\esper(\Theta_{i,j})\,$ and $\,\sigma_{i,j}^2:=\vari(\Theta_{i,j}),$ then as a consequence of the linearity property for expected values it is immediate to obtain:
\begin{equation}\label{eq:esperanza}
  \mu_j \,:=\, \esper(\Theta_j) \,=\,\sum_{i\,=\,1}^N\beta_i\mu_{i,j}\,,\qquad j\in\{1,\ldots,J\},
\end{equation}
and therefore the expectation of $\Theta_j$ only depends on the \textit{marginal} expectations of the random variables $\,\{\Theta_{i,j}\,:\,i=1,\ldots,N\}\,$ with no consideration needed about the possible dependence among them (dependence-free). However, in calculating the variance of $\Theta_j$ the possible dependence among them cannot be ignored since pairwise stratum covariances are required for each candidate $j$:
\begin{eqnarray}
  \sigma_j^2 := \vari(\Theta_j) &=& \sum_{i\,=\,1}^N\beta_i^2\sigma_{i,j}^2 \,+\, 2\!\sum_{1\,\leq\,i\,<}\sum_{r\,\leq\,N}\beta_i\beta_r\cov(\Theta_{i,j},\Theta_{r,j}) \label{eq:varianza} \\
	                              &=& \sum_{i\,=\,1}^N\sum_{r\,=\,1}^N\beta_i\beta_r\cov(\Theta_{i,j},\Theta_{r,j}) \nonumber \\
	                              &\leq& \sum_{i\,=\,1}^N\sum_{r\,=\,1}^N\beta_i\beta_r\sigma_{i,j}\sigma_{r,j} = \bigg(\sum_{i\,=\,1}^N\beta_i\sigma_{i,j}\bigg)^2 \label{eq:varianzaMax}
\end{eqnarray}
where (\ref{eq:varianzaMax}) is an immediate consequence of the \textit{Cauchy-Schwarz inequality,} see for example Casella and Berger (2002), and provides a maximum possible value for the variance $\,\sigma_j^2.$ Let $\,I_j(\alpha)\,$ be an interval of probability $\,0<\alpha<1\,$ for $\,\Theta_j,$ that is $\,\prob[\Theta_j\in I_j(\alpha)]=\alpha.$ Suppose there exists a value $\,z_{\alpha}\,$ such that:
\begin{equation}\label{eq:intervalo}
   I_j(\alpha) \,=\, [\,\mu_j - z_{\alpha}\sigma_j\,,\,\mu_j + z_{\alpha}\sigma_j\,]
\end{equation}
and the analogous ones for $\,\Theta_{i,j}\,$ through $I_{i,j}(\alpha)=[\mu_{i,j}-z_{\alpha}\sigma_{i,j},\mu_{i,j}+z_{\alpha}\sigma_{i,j}].$ If the interval $I_j^{*}(\alpha)$ is a convex linear combination of the intervals $\,I_{i,j}(\alpha)\,$ with weights $\,\beta_i=n_i/n\,$ then:
\begin{eqnarray}
  I_j^{*}(\alpha) &=& \sum_{i\,=\,1}^N\beta_i I_{i,j}(\alpha) \nonumber \\
	                &=& \bigg[\,\sum_{i\,=\,1}^N\beta_i\mu_{i,j} - z_{\alpha}\sum_{i\,=\,1}^N\beta_i\sigma_{i,j}\,\,,\,\,\sum_{i\,=\,1}^N\beta_i\mu_{i,j} + z_{\alpha}\sum_{i\,=\,1}^N\beta_i\sigma_{i,j}\,\bigg] \nonumber \\
									&=& [\,\mu_j-z_{\alpha}\sigma_j^{*}\,,\,\mu_j+z_{\alpha}\sigma_j^{*}\,] \label{eq:intervaloMax}
\end{eqnarray}
where the standard deviation$\,\sigma_j^{*}=\sum_{i=1}^N\beta_i\sigma_{i,j}\,$ is the maximum possible value for $\,\sigma_j\,$ accordingly to (\ref{eq:varianzaMax}). In case the assumption of independence between strata is correct then $\,\cov(\Theta_{i,j},\Theta_{r,j})=0\,$ for all $i\neq r$ and, as a consequence of (\ref{eq:varianza}), the standard deviation would be $\,\sigma_j^{\perp}=\sqrt{\sum_{i\,=\,1}^N\beta_i^2\sigma_{i,j}^2}\,,$ a value that is smaller than $\,\sigma_j^{*}\,$ and therefore would lead to a narrower interval $\,I_j^{\perp}(\alpha)=[\mu_j-z_{\alpha}\sigma_j^{\perp},\mu_j+z_{\alpha}\sigma_j^{\perp}]\subset I_j^{*}(\alpha).$ But if the independence assumption between strata is not correct, the resulting interval $\,I_j^{\perp}\,$ could be narrower than it should, and therefore with a probability lower than the desired $\,\alpha\,$ level. In the other hand, using interval $\,I_j^*\,$ assuming maximum covariances between strata could lead to wider intervals with a probability greater than the desired $\,\alpha\,$ level: $I_j^{\perp}(\alpha)\subset I_j(\alpha)\subset I_j^{*}(\alpha).$

\bigskip

\noindent One would like to obtain $\,I_j(\alpha)\,$ as in (\ref{eq:intervalo}) with the correct value of $\,\sigma_j\,$ through (\ref{eq:varianza}) but this would require to estimate $\,N(N-1)/2\,$ covariances, which in turn require information that is usually not available in this particular matter (elections). Assuming that the value of $\,\sigma_j\,$ belongs to the interval $\,[\,\sigma_j^{\perp}\,,\,\sigma_j^*\,]\,$ then a value $0\leq\delta_j\leq 1$ should exist such that
\begin{equation}\label{eq:delta}
  \sigma_j \,=\, (1-\delta_j)\sigma_j^{\perp} \,+\, \delta_j\sigma_j^{*}
\end{equation}
and from this a probability $\,\alpha\,$ interval for $\,\Theta_j\,$ would be:
\begin{eqnarray}
  I_j(\alpha) &=& \mu_j \,\pm\, z_{\alpha}[(1-\delta_j)\sigma_j^{\perp} \,+\, \delta_j\sigma_j^*] \nonumber \\
	            &=& (1-\delta_j)\mu_j \,+\, \delta_j\mu_j \,\pm\, z_{\alpha}(1-\delta_j)\sigma_j^{\perp} \,\pm\, z_{\alpha}\delta_j\sigma_j^* \nonumber \\
							&=& (1-\delta_j)I_j^{\perp}(\alpha) \,+\, \delta_jI_j^*(\alpha)\,. \label{eq:intervalo2}
\end{eqnarray}

\noindent In (\ref{eq:intervalo2}) still there is the problem of estimating $\,\delta_j\,$ for each candidate $\,j\in\{1,\ldots,J\}.$ To have an idea of usual values for $\delta_j$ one could estimate such values from past or similar elections through the following:

\bigskip\medskip

\noindent \underline{\textbf{Algorithm 1}}
\begin{enumerate}
  \item Estimate intervals $\,I_j^{\perp}(\alpha)\,$ and $\,I_j^*(\alpha)$ for each $j\in\{1,\ldots,J\}.$
	\item Simulate a large amount of stratified samples.
	\item Determine each $\,\delta_j\,$ such that the interval (\ref{eq:intervalo2}) contains the true value $\,\theta_j\,$ in $100\alpha\%$ of the simulations (coverage probability $\alpha$) from step 2.
\end{enumerate}

\medskip

\noindent In summary, the main idea of this section was to show that an incorrect assessment of the degree of dependence (or lack thereof) between strata may lead to narrower or wider intervals that could have a different coverage probability level than the desired one, and therefore some adjustment has to be made to take such dependence into consideration. In the particular case where the strata are electoral districts, one may intuitively anticipate there could be some degree of positive dependence: an overall better/worse performance for a candidate is possibly due to better/worse performance in many districts simultaneously.

\section{A copula based model}

\noindent The \textit{probability of victory} for each registered candidate involves calculations through an appropriate transformation of the $J-\text{variate}$ random vector $\,(\Theta_1,\ldots,\Theta_J).$ If $\,\mathcal{W}_j\,$ represents the event that registered candidate $j\in\{1,\ldots,J-3\}$ wins, then:
\begin{equation}\label{eq:probVictoria}
  \prob(\mathcal{W}_j) \,=\, \prob(\max\{\Theta_1,\ldots,\Theta_{J-3}\}=\Theta_j)\,,
\end{equation}
where $\,0<\Theta_j<1\,$ and $\,\sum_{j=1}^J\Theta_j=1\,$ as a consequence of (\ref{eq:propVotos}), which in turn implies that, for example, $\Theta_J=1-\sum_{j=1}^{J-1}\Theta_j$ and therefore it is just necessary to have the joint probability distribution of the $(J-1)-\text{variate}$ random vector $\,(\Theta_1,\ldots,\Theta_{J-1})\,$ with \textit{support} in the set known as $(J-1)-\text{dimensional}$ \textit{simplex} defined by:
\begin{equation}\label{eq:simplex}
  \mathcal{S}:=\{(\theta_1,\ldots,\theta_{J-1}):\theta_1>0,\ldots,\theta_{J-1}>0\,;\,\theta_1+\cdots+\theta_{J-1}<1\}.
\end{equation}

\medskip

\noindent Let $\Theta_j$ be a continuous random variable, for all $j,$ with marginal distribution function $F_j$ and support the open interval $\,]0,1[\,,$ then as a result of the outstanding theorem by Sklar (1959), see Nelsen (2006), there exists a unique functional link $\mathbf{C}$ (known as \textit{copula} function) between the joint probability distribution $\mathbf{H}$ of $\,(\Theta_1,\ldots,\Theta_{J-1})\,$ and its univariate marginal distributions, that is:
\begin{equation}\label{eq:Sklar}
  \mathbf{H}(\theta_1,\ldots,\theta_{J-1})\,=\,\mathbf{C}\big(F_1(\theta_1),\ldots,F_{J-1}(\theta_{J-1})\big).
\end{equation}
Instead of dealing with high-dimensional multivariate models, with the estimation challenges due to the so called \textit{curse of dimensionality}, it would be enough to obtain pairwise bivariate joint distributions:
\begin{equation}\label{eq:Sklar2D}
  \mathbf{H}_{j,\ell}(x,y)\,=\,\mathbf{C}_{j,\ell}\big(F_j(x),F_{\ell}(y)\big)\,,\qquad j\neq\ell,
\end{equation}
and calculate the pairwise probabilities $\prob(\Theta_j>\Theta_{\ell})$ for $j\neq\ell$ with $j$ and $\ell$ in $\{1,\ldots,J-3\}.$ Therefore, the marginal distributions $F_j$ and the bivariate copulas $\mathbf{C}_{j,\ell}$ have to be estimated from the stratified sample information described in section 2.

\bigskip

\noindent For each candidate $j$ in stratum $i$ there will be information from $c_i$ polling stations in the stratified sample, with observed values $\{x_{i,j}^{(1)}, x_{i,j}^{(2)},\ldots,x_{i,j}^{(c_i)}\}$ from which it would be possible to calculate the observed proportions of votes at each polling station in the sample:
\begin{equation}\label{obsthetas}
  \theta_{i,j}^{(k)}\,=\,\,\frac{x_{i,j}^{(k)}}{n_i^{(k)}}\,\,,\qquad k\in\{1,\ldots,c_i\},
\end{equation}
where $n_i^{(k)}$ is the number of potential voters in polling station $k.$ From data (\ref{obsthetas}) it is possible to estimate (in a parametric or non-parametric fashion) each $F_{i,j},$ that is the marginal distribution function for $\Theta_{i,j},$ and therefore $I_{i,j}(\alpha),$ $\mu_{i,j}$ and $\sigma_{i,j}^2.$ Moreover, as a consequence of (\ref{eq:esperanza}), it is possible to directly estimate the overall means for each candidate $j:$
\begin{equation}\label{medias}
   \mu_j \,=\, \esper(\Theta_j) \,=\, \sum_{i\,=\,1}^N \frac{n_i}{n}\,\mu_{i,j}\,,
\end{equation}
but for the overall intervals $I_j(\alpha)$ and the probability of victory some extra work needs to be done, because variances and covariances require calculations regarding dependence between strata and between candidates.

\medskip

\noindent As a consequence of what has already been discussed in section 3, under zero and maximum pairwise covariances $\cov(\Theta_{i,j},\Theta_{r,j})$ between strata ($i\neq r$), for each candidate $j$ the marginal variances are obtained, respectively:
\begin{equation}\label{varmarginales}
  \sigma_j^{\perp} \,=\, \sqrt{\sum_{i\,=\,1}^N\Big(\frac{n_i}{n}\Big)^2\,\sigma_{i,j}^2}\quad < \quad\sum_{i\,=\,1}^N \frac{n_i}{n}\,\sigma_{i,j}\,=\,\sigma_j^*\,.
\end{equation}

\noindent With (\ref{varmarginales}) along with $\delta_j$ values obtained by Algorithm 1 in section 3, it is obtained for each candidate $j$ the following overall marginal standard deviation for each candidate $j$:
\begin{equation}\label{varmarginal}
  \sigma_j \,=\, (1-\delta_j)\sigma_j^{\perp} \,+\, \delta_j\sigma_j^*\,.
\end{equation}

\medskip

\noindent In a parametric fashion, it would be possible to estimate each marginal distribution $F_j$ as follows. Consider $X_{i,j}^{(k)}$ as a Binomial random variable with known parameter $n_i^{(k)}$ and unknown parameter $\theta_{i,j}\,.$ Then the total number of votes for candidate $j$ in stratum $i$ from a sample of size $c_i,$  that is $X_{i,j}=\sum_{k=1}^{c_i}X_{i,j}^{(k)},$ would also be a Binomial random variable with known parameter $\sum_{k=1}^{c_i}n_i^{(k)}$ and unknown parameter $\theta_{i,j}\,.$ Given observed sample values $\{x_{i,j}^{(1)}, x_{i,j}^{(2)},\ldots,x_{i,j}^{(c_i)}\}$ and adopting a bayesian approach with a non-informative conjugate prior Beta distribution (see for example Bernardo and Smith, 1994) it is obtained as a posterior distribution for $\theta_{i,j}$ a Beta distribution with parameters:
\begin{equation}\label{betaparam}
  \alpha_{i,j}\,=\,\frac{1}{2}\,+\sum_{k\,=\,1}^{c_i}x_{i,j}^{(k)} \qquad\text{and}\qquad \beta_{i,j}\,=\,\frac{1}{2}\,+\sum_{k\,=\,1}^{c_i}[\,n_i^{(k)} - x_{i,j}^{(k)}\,]\,,
\end{equation}
which determine $F_{i,j}\,,$ and in particular:
\begin{equation}\label{mediavarianza}
  \mu_{i,j}\,=\,\frac{\alpha_{i,j}}{\alpha_{i,j}+\beta_{i,j}}\,=\,\frac{\frac{1}{2}+\sum_{k=1}^{c_i}x_{i,j}^{(k)}}{1+\sum_{k=1}^{c_i}n_i^{(k)}} \qquad\text{and}\qquad \sigma_{i,j}^2\,=\,\frac{\alpha_{i,j}\beta_{i,j}}{(\alpha_{i,j}+\beta_{i,j})^2(\alpha_{i,j}+\beta_{i,j}+1)}\,.
\end{equation}

\noindent Finally, each $F_j$ may be approximated with a Beta distribution with mean $\mu_j$ as in (\ref{medias}) and variance $\sigma_j^2$ as in (\ref{varmarginal}), and thereafter obtaining the overall intervals $I_j(\alpha)$ is straightforward.

\bigskip

\noindent To calculate the pairwise probabilities $\prob(\Theta_j>\Theta_{\ell})$, the above marginal distributions $F_j$ are needed but not enough for such purpose, since from (\ref{eq:Sklar2D}) a copula estimation is required. Pairwise overall covariances between candidates $j\neq\ell$ are obtained by:
\begin{eqnarray}
  \cov(\Theta_j\,,\Theta_{\ell}) &=& \cov\bigg(\sum_{i\,=\,1}^N\frac{n_i}{n}\,\Theta_{i,j}\,\,,\,\sum_{r\,=\,1}^N\frac{n_r}{n}\,\Theta_{r,l}\bigg)\,, \nonumber \\
	&=& \frac{1}{n^2}\sum_{i\,=\,1}^N\sum_{r\,=\,1}^N n_i n_r \cov(\Theta_{i,j}\,,\Theta_{r,\ell})\,, \label{cov1} \\
	&=& \frac{1}{n^2}\sum_{i\,=\,1}^N n_i^2 \cov(\Theta_{i,j}\,,\Theta_{i,\ell}) \,+\, \frac{1}{n^2}\sum\sum_{\!\!\!\!\!\!\!\!i\,\neq\,r} n_i n_r \cov(\Theta_{i,j}\,,\Theta_{r,\ell})\,, \label{cov2}
\end{eqnarray}
where same stratum covariances between candidates $\cov(\Theta_{i,j}\,,\Theta_{i,\ell})$ in the right first term of (\ref{cov2}) may be estimated from pairwise observations $\{(\theta_{i,j}^{(k)},\theta_{i,\ell}^{(k)}):k=1,\ldots,c_i\}$ from (\ref{obsthetas}), but for different stratum covariances between candidates as in the right second term of (\ref{cov2}) there will be no paired information available, therefore some sort of \textit{simplifying assumption} is needed: $\cov(\Theta_{i,j}\,,\Theta_{r,\ell})\approx\cov(\Theta_{i,j},\Theta_{i,\ell}),$ which when substituted in (\ref{cov1}) leads to:
\begin{equation}\label{covarianzaSA}
  \cov(\Theta_j\,,\Theta_{\ell}) \,=\, \sum_{i\,=\,1}^N\frac{n_i}{n}\,\cov(\Theta_{i,j},\Theta_{i,\ell})\,,
\end{equation}
and therefore the candidates' pairwise Pearson correlations are given by:
\begin{equation}\label{corrPearson}
  \rho_{j,\ell}\,=\,\,\frac{\cov(\Theta_j\,,\Theta_{\ell})}{\sqrt{\vari(\Theta_j)\vari(\Theta_{\ell}))}} \,=\, \sum_{i\,=\,1}^N\frac{n_i}{n}\,corr(\Theta_{i,j},\Theta_{i,\ell}) \,.
\end{equation}

\medskip

\noindent A popular bivariate distribution where all the dependence is determined by Pearson's correlation coefficient is the bivariate \textit{Normal} (or \textit{Gaussian}) distribution. This distribution has the disadvantage that its univariate marginals have to be also Normally distributed, and since marginally the proportions of interest $0<\Theta_j<1$ this is an inconvenience that may be overcome using the \textit{Normal} (or \textit{Gaussian}) copula $\mathbf{C}_{\rho}$ instead, see for example Salvadori \textit{et al.}(2007), and combine it with the marginals $F_j$ applying Sklar's Theorem as in (\ref{eq:Sklar2D}). The Gaussian copula does not have a closed form, therefore the calculation of pairwise probabilities $\prob(\Theta_j>\Theta_{\ell})$ for $j\neq\ell$ with $j$ and $\ell$ in $\{1,\ldots,J-3\}$ needs to be done through simulation, accordingly with the following:

\bigskip\medskip

\noindent \underline{\textbf{Algorithm 2}}
\begin{enumerate}
  \item Simulate a large number $m$ of bivariate observations $(u_t,v_t)$ from a Gaussian copula with correlation parameter $\rho_{j,\ell}$ obtained by (\ref{corrPearson}), with $j,\ell\in\{1,\ldots,J-3\},$ $j<\ell.$
	\item Calculate $\{(\hat{\theta}_j^{(t)},\hat{\theta}_{\ell}^{(t)})=(F_j^{-1}(u_t),F_{\ell}^{-1}(v_t)):t=1,\ldots,m\}.$
	\item Estimate $\prob(\Theta_j>\Theta_{\ell})\,\approx\,\frac{1}{m}\sum_{t=1}^m \indic\{\hat{\theta}_j^{(t)}>\hat{\theta}_{\ell}^{(t)}\}.$
\end{enumerate}

\bigskip

\noindent The proportion of votes $\Theta_j$ is with respect to all the $n$ potential voters, but it is also of interest the proportion of votes with respect to the total \textit{votes effectively cast} at the polling stations (that is to say, without considering abstentions), which may be obtained as:
\begin{equation}\label{eq:propVotosEfec}
  \Lambda_j \,=\, \frac{X_j}{\sum_{\ell=1}^{J-1}X_{\ell}} \,=\, \frac{\Theta_j}{\sum_{\ell=1}^{J-1}\Theta_{\ell}} \,=\, \frac{\Theta_j}{1-\Theta_J} \,,\qquad j\in\{1,\ldots,J-1\},
\end{equation}
where $1-\Theta_J$ is the proportion of \textit{effective participation} in the election, that is the complement of abstentionism. Transformation (\ref{eq:propVotosEfec}) has no impact in the probability of victory since clearly $\prob(\Theta_j>\Theta_{\ell})=\prob(\Lambda_j>\Lambda_{\ell}),$ but it does for (marginal) point and interval estimation. This may be addressed estimating the bivariate joint distribution of $(\Theta_j,1-\Theta_J)$ for each $j\in\{1,\ldots,J-1\},$ again by (\ref{eq:Sklar2D}), and then (through simulation as in Algorithm 2 but in step 3 calculating $\hat{\lambda}_j^{(t)}=\hat{\theta}_j^{(t)}/(1-\hat{\theta}_J^{(t)})$) estimating the probability distribution $F_{\Lambda_j}\,,$ from which it is possible to obtain point and interval estimations through $\,\esper(\Lambda_j)\,$ and $\,[\,a_j\,,\,b_j\,],$ respectively, with $\,a_j<b_j\,$ such that $\,F_{\Lambda_j}(b_j)-F_{\Lambda_j}(a_j)=\gamma\,$ and minimizing interval length $\,b_j-\,a_j\,,$ for a desired probability level $\,0<\gamma<1\,:$

\bigskip\medskip

\noindent \underline{\textbf{Algorithm 3}}
\begin{enumerate}
  \item Simulate a large number $m$ of bivariate observations $(u_t,v_t)$ from a Gaussian copula with correlation parameter $\rho_{j,J}$ obtained by (\ref{corrPearson}), with $j\in\{1,\ldots,J-1\}.$
	\item Calculate $\{(\hat{\theta}_j^{(t)},\hat{\theta}_{J}^{(t)})=(F_j^{-1}(u_t),F_{J}^{-1}(v_t)):t=1,\ldots,m\}.$
	\item Calculate $\{\hat{\lambda}_j^{(t)}=\hat{\theta}_j^{(t)}/(1-\hat{\theta}_J^{(t)}):t=1,\ldots,m\}.$
	\item Interval estimation through empirical quantiles of $\hat{\lambda}_j^{(t)}:\,[\,a_j\,,\,b_j\,]\,$ such that $\,F_{\Lambda_j}(b_j)-F_{\Lambda_j}(a_j)=\gamma\,$ and minimizing interval length $\,b_j-\,a_j\,,$ for a desired probability level $\,0<\gamma<1.$
\end{enumerate}

\bigskip

\noindent In summary, the main idea of this section was to provide marginal interval estimations $I_j(\alpha)$ for each candidate $j$ that take into consideration dependence between strata, to estimate pairwise probabilities $\prob(\Theta_j>\Theta_{\ell})$ with a copula function approach that takes into consideration dependence between candidates and between strata, and marginal interval estimations for the proportion of votes with respect to the total votes effectively cast at the polling stations, which need to consider both types of dependence.

\section{Mexico's 2006, 2012, and 2018 presidential elections}

\noindent The proposed copula based model is now applied to data from Mexico's presidential elections in the years 2006, 2012, and 2018, published by the \textit{National Electoral Institute} (INE: 2006, 2012, 2018). In the official quick counts, the basis for stratification was the 300 federal electoral districts, with a refinement in 2006 and 2012 to more than 480 strata by splitting each district into urban/non-urban polling stations, and with 350 strata in 2018, without urban/non-urban refinement, but with more than 300 strata because in 10 of 32 states local districts (instead of federal districts) were used due to concurrent governor elections. In the present study, for the sake of comparability, the 300 federal districts will be used as strata for the analysis of the three elections. 

\bigskip

\noindent Recall that the $\delta_j$ values as in (\ref{eq:delta}) and (\ref{eq:intervalo2}) are needed to include the dependence effect among strata for interval estimation. Such values may be estimated \textit{ex post} through Algorithm 1, see Table \ref{DELTAS} where estimations where made for the three leading candidates in the three elections: PAN (\textit{National Action Party}, alone in 2006 and 2012, but leading a coalition of political parties in 2018), PRI (\textit{Revolutionary Institutional Party}, leading different coalitions of political parties in each election), and AMLO (Andr\'es Manuel L\'opez Obrador, candidate for coalitions of political parties leaded by the PRD \textit{--Democratic Revolution Party--} in 2006 and 2012, but a candidate in 2018 for a coalition of political parties leaded by the recently created \textit{National Regeneration Movement} -- MORENA). Even though the planned sample size was quite similar for the three elections, the effectively received stratified sample sizes were quite different: 7263, 6260, and 5254, respectively for 2006, 2012, and 2018, so the same sample sizes are used in the following simulation study.

\bigskip
\begin{table}[h]
  \begin{center}
    \includegraphics[width = 7cm, keepaspectratio]{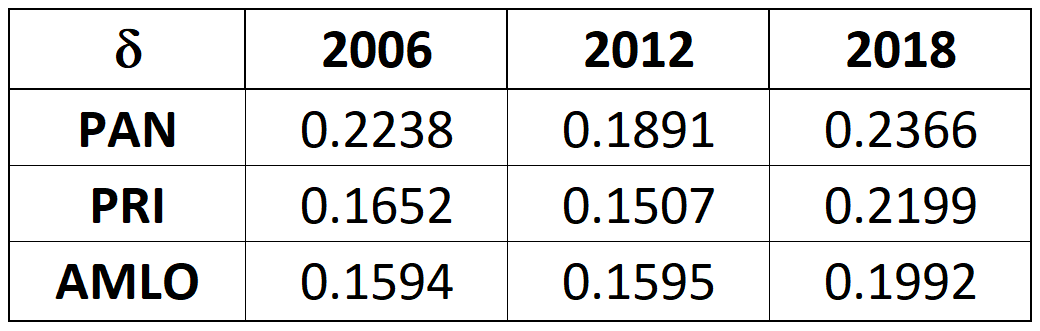}
  \end{center}
  \caption{Estimated $\delta_j$ values through Algorithm 1 for a 95\% coverage probability in Mexico's presidential elections in the years 2006, 2012, and 2018, drawing 10000 stratified samples of the size used in each election (for the proportion of votes with respect to all the potential voters).}
  \label{DELTAS}
\end{table}
\bigskip

\noindent The results of the proposed copula based model are compared to what is obtained applying the model by Mendoza and Nieto-Barajas (2016), which is summarized as follows:
\begin{equation}\label{Mendoza1}
  X_{i,j}^{(k)}\,|\,\theta_{i,j},\tau_{i,j}\,\sim\,\text{Normal}\Big(\,n_i^{(k)}\theta_{i,j}\,,\,\frac{\tau_{i,j}}{n_i^{(k)}}\,\Big)
\end{equation}
with $\tau_{i,j}/n_i^{(k)}$ being the precision parameter where, for each candidate, $\tau_{i,j}$ is assumed to be constant within the corresponding stratum (and unrelated to $\theta_{i,j}$). Moreover, they assume that $X_{i,j}^{(k)}$ is independent of $X_{i,j'}^{(k)},$ for $j\neq j'.$ As the authors mention: \textsl{``This is perhaps a stronger assumption, however, previous analysis with a more complex model that assumed dependence between these variables showed that the dependence was too weak and could be disregarded.''} They argue that the dependence between the parameters of interest (the $\lambda$'s) is recovered simulating bayesian posterior values for $\theta_{i,j}$ and transforming such values through formulas (\ref{eq:propVotos}) and (\ref{eq:propVotosEfec}). After combining a non-informative prior for $(\theta_{i,j},\tau_{i,j})$ with information from the data (received stratified sample) they obtain a posterior distribution conditional on the data that is proportional to the product of a truncated normal distribution for $\theta_{i,j},$ conditional on $\tau_{i,j},$ and a gamma distribution for $\tau_{i,j},$ that is:
\begin{equation}\label{Mendoza2}
\begin{split}
  p(\theta_{i,j},\tau_{i,j})\,\propto\,Normal\bigg(\theta_{i,j}\,\bigg|\,\frac{\sum_{k=1}^{c_i}x_{i,j}^{(k)}}{\sum_{k=1}^{c_i}n_i^{(k)}}\,\,,\,\,\tau_{i,j}\sum_{k=1}^{c_i}n_i^{(k)}\bigg)I(0<\theta_{i,j}<1) \\ \times\, Gamma\bigg(\tau_{i,j}\,\bigg|\,\frac{c_i-1}{2}\,\,,\,\,\frac{1}{2}\bigg\{\sum_{k=1}^{c_i}\frac{(x_{i,j}^{(k)})^2}{n_i^{(k)}}\,-\,\frac{\big(\sum_{k=1}^{c_i}x_{i,j}^{(k)}\big)^2}{\sum_{k=1}^{c_i}n_i^{(k)}}\bigg\}\bigg)
\end{split}
\end{equation}
The authors clarify that in (\ref{Mendoza2}) the posterior distribution for $\tau_{i,j}$ is proper only if the sample number of polls (in stratum $i$) satisfies $c_i\geq 2.$ They missed to mention that it is also required that $x_{i,j}\neq 0$ otherwise the second parameter of the gamma distribution in (\ref{Mendoza2}) would be equal to zero, which is non admissible for such distribution. Such (typically non crucial) restrictions are not needed under the proposed copula based model.

\bigskip

\noindent Under the proposed copula based model, and for each of the three elections analyzed (2006, 2012, and 2018), the probability of victory is estimated through Algorithm 2, calculating for the leading candidate the paired probability of winning over the second leading one, and for the others the paired of probability of winning over the leading candidate. Interval estimations for the proportion of votes with respect to the total \textit{votes effectively cast} at the polling stations, for the three leading candidates, are obtained through Algorithm 3, under 10000 simulated stratified samples from the official total vote count, along with an estimation of the \textit{coverage probability} of the estimated intervals, that is, the proportion of simulations where the estimated intervals contain the target value: the proportion of total votes effectively cast at the polling stations for each candidate in the final total count. These results are compared to the ones obtained by applying the model proposed by Mendoza and Nieto-Barajas (2016). All the calculations and simulations were performed in the programming language by R Core Team (2018) and copula calculations with the R package \texttt{copula} by Hofert \textit{et al.}(2017).

\bigskip
\begin{table} 
  \begin{center}
    \includegraphics[width = 17cm, keepaspectratio]{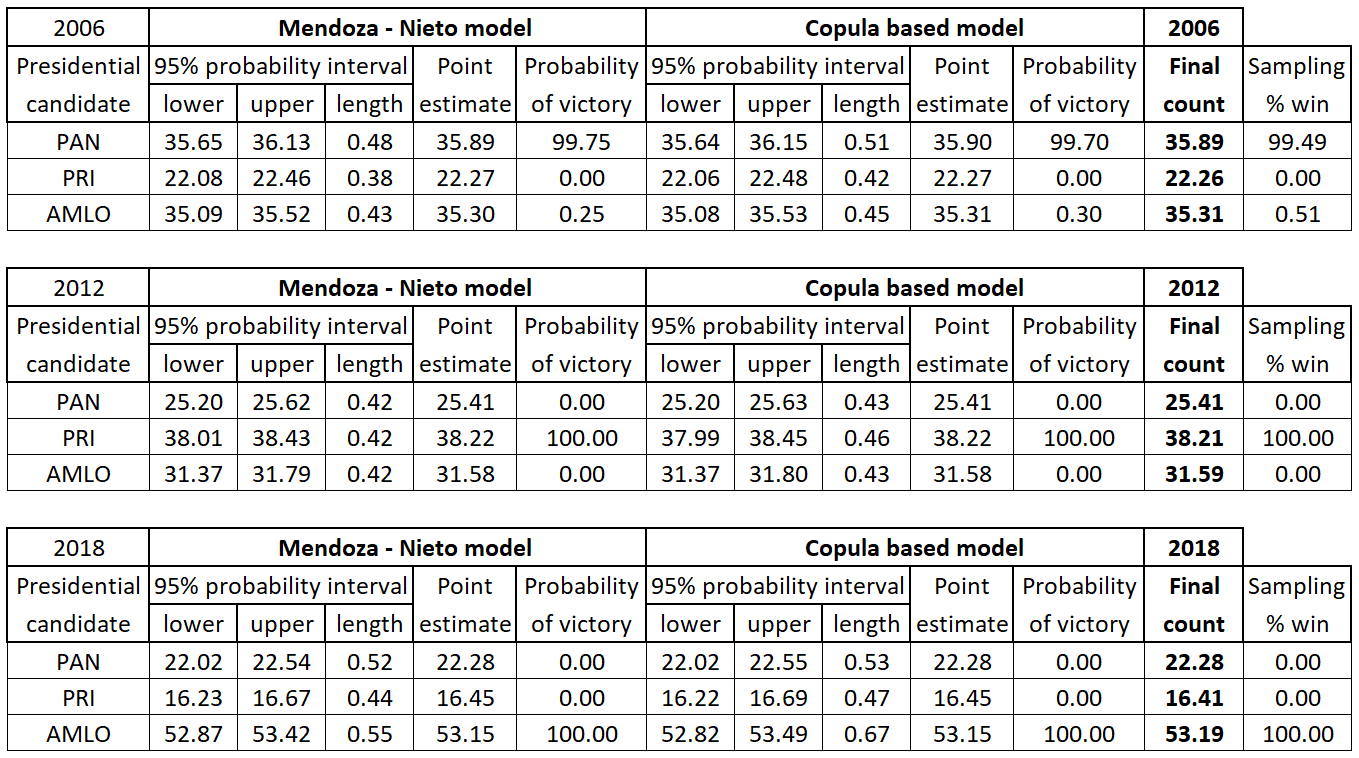}
  \end{center}
  \caption{Average of 10000 interval and point estimations (for the proportion of votes with respect to the total \textit{votes effectively cast} at the polling stations), and probability of victory, for each candidate and election. All the amounts are in percentage points.}
  \label{INTERVALOS}
\end{table}
\bigskip

\noindent In Table \ref{INTERVALOS} it is summarized the average 95\% probability intervals obtained in the simulations under the two methods under comparison, along with their estimated probabilities of victory for each candidate, and compared to the results in the total final count of votes. All the amounts are in percentage points. The last column is the percentage of simulated stratified samples where each candidate turned up to get the highest percentage of all according to formula (\ref{estimadorderazon}), which is to be compared to the probability of victory estimated by each model. For both models, the average 95\% probability intervals appear to be fairly centered around the target value (percentage of votes effectively cast in the final count), but in all cases the interval lengths under the copula based model are slightly larger than the interval lengths under the Mendoza-Nieto model. So to decide where each model could be under or over estimating such length, the empirical coverage probability of the intervals is estimated under each method for each candidate and election, which ideally should be also 95\%, so the closer the better, see Table \ref{COBERTURA}.

\bigskip
\begin{table} 
  \begin{center}
    \includegraphics[width = 10cm, keepaspectratio]{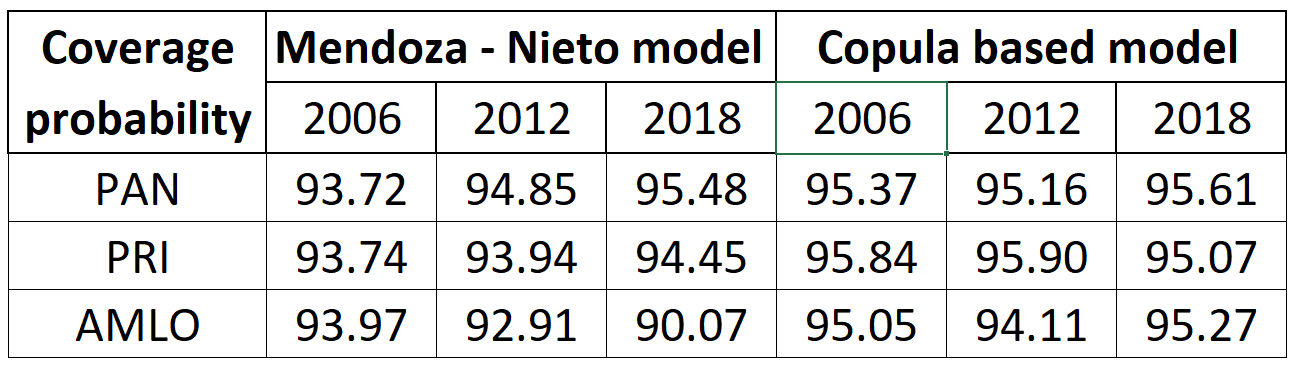}
  \end{center}
  \caption{Coverage probability (in percentage) of the 95\% probability intervals (for the proportion of votes with respect to the total \textit{votes effectively cast} at the polling stations) estimated under the two methods under comparison (10000 simulations).}
  \label{COBERTURA}
\end{table}
\bigskip

\noindent In all cases, the coverage probability under the Mendoza-Nieto model is lower than under the proposed copula model, in many cases below the ideal 95\%, which is consistent with narrower intervals as summarized in Table \ref{INTERVALOS}, and provides empirical evidence that confirms what theoretically is argued in section 3: the need of considering appropriate dependence among strata for the case of interval estimation, which is not considered in Mendoza-Nieto's model.

\bigskip

\noindent Regarding the probability of victory, in 2012 and 2018 elections the difference between the first and second place was large enough (6.62 and 30.91 percentage points, respectively) so that under no simulated scenario was possible to observe something different. It was just in 2006 election where the difference of 0.58 percentage points between the first and second place was small enough so that in a few simulated scenarios a different result was obtained, and similar probabilities of victory are obtained under both methods. Recall that in section 4 it has been argued theoretically that dependence between candidates has an impact on the calculation of pairwise probabilities of victory, and that Mendoza and Nieto-Barajas (2016) argued that under a previous analysis they made (without disclosing it) with a model that assumed dependence between candidates showed that the dependence was too weak and could be disregarded, which seems to be the case for the particular election of 2006, but the possibility this might no be the case in some other election is open, and the proposed copula based model is prepared to account for it if necessary.

\section{Final remarks}

	
\noindent The main contribution of the present work is to prove that an incorrect assessment of the degree of dependence (or lack thereof) between strata and between candidates may lead to narrower or wider intervals that could have a different coverage probability level than the desired one, and therefore some adjustment has to be made to take such dependencies into consideration. Moreover, a copula based model is proposed for calculating such interval estimations, and for the purpose of estimating pairwise probabilities of victory between candidates, that also need to take into consideration both types of dependence.

\bigskip
	
\noindent A simulation study was performed to compare the proposed copula based model versus the model by Mendoza and Nieto-Barajas (2016), making use of official data from Mexico's presidential elections in the years 2006, 2012, and 2018. For such elections it became clear that taking into consideration dependence between strata and between candidates was necessary to achieve coverage probabilities close to a desired level of 95\%. For the probability of victory, although theoretically it is proved that dependence between candidates plays a role in its calculation, in such particular examples was not significantly relevant for such purpose, but the proposed model is ready to account for it whenever necessary. 

\bigskip

\noindent 


\noindent

\section*{Acknowledgement}

\noindent This work was partially supported by Programa UNAM--DGAPA--PAPIIT project \textbf{IN115817}.

\section*{References}




\noindent Bernardo, J.M., Smith, A.F.M. (1994) \textit{Bayesian Theory.} Wiley (New York).\medskip


\noindent Casella, G.,  Berger, R.L. (2002) \textit{Statistical Inference.} Duxbury (Pacific Grove).\medskip







\noindent Hofert, M., Kojadinovic, I., Maechler, M., Yan, J. (2017) \texttt{copula}: \textit{Multivariate Dependence with Copulas.} R   package version 0.999-18 URL https://CRAN.R-project.org/package=copula\medskip
	

\noindent INE: Instituto Nacional Electoral (2006, 2012) \textit{Bases de Datos de los Procesos Electorales Locales en M\'exico.} URL http://siceef.ine.mx/downloadDB.html \medskip

\noindent INE: Instituto Nacional Electoral (2018) \textit{Conteos R\'apidos: Procesos Electorales Federal y Locales 2017-2018} (M\'exico). URL https://www.ine.mx/conteos-rapidos-procesos-electorales-federal-locales-2017-2018 \medskip

\noindent Lindley, D.V. (2000) The Philosophy of Statistics. \textit{Journal of the Royal Statistical Society. Series D (The Statistician)} \textbf{49} (3), 293--337.\medskip

\noindent Mendoza, M., Nieto-Barajas, L.E. (2016) Quick counts in the Mexican presidential elections: A Bayesian approach. \textit{Electoral Studies} \textbf{43}, 124--132.\medskip



\noindent Nelsen, R.B. (2006) \textit{An Introduction to Copulas.} Springer.\medskip


\noindent R Core Team (2018) R: A language and environment for statistical computing. \textit{R Foundation for Statistical Computing},   Vienna, Austria. URL https://www.R-project.org/\medskip

\noindent Salvadori, G., De Michele, C., Kottegoda, N.T., Rosso, R. (2007) \textit{Extremes in Nature. An Approach Using Copulas.} Springer.\medskip
	
\noindent S\"{a}rndal, C.-E., Swensson, B., Wretman, J. (1992) \textit{Model Assisted Survey Sampling.} Springer.\medskip

\noindent Sklar, A. (1959) Fonctions de r\'epartition \`a $n$ dimensions et leurs marges. \textit{Publ. Inst. Statist. Univ. Paris,} \textbf{8}, 229--231.\medskip


\end{document}